\documentclass[a4paper,11pt]{article}
\usepackage{pos}

\usepackage{amsmath}
\usepackage{amsfonts}
\usepackage{physics}
\usepackage{slashed}
\usepackage{simplewick}

\usepackage{graphicx}
\usepackage{float}
\usepackage[caption=false]{subfig}

\usepackage{tikz}
\usetikzlibrary{calc}
\usetikzlibrary{patterns}
\usetikzlibrary{positioning,decorations.pathmorphing,decorations.markings}
\usetikzlibrary{shapes,arrows}

\usepackage{feynmp-auto}
\usepackage{multirow}
\usepackage{adjustbox}

\usepackage{enumitem}

\usepackage{hyperref}

\usepackage{color}
\usepackage{comment}
\def\str#1{{\it str} \{ #1 \}}

\def\ket#1{| #1 \ra }

\def\la{\langle}
\def\ra{\rangle}

\def\l{\left}
\def\r{\right}

\def\beq{\begin{equation}}
\def\eeq{\end{equation}}
\def\bea{\begin{eqnarray}}
\def\eea{\end{eqnarray}}
\def\overliner{\begin{array}}
\def\earr{\end{array}}

\def\sec#1{{sez.~\ref{#1}}}

\def\mev{\mbox{ MeV}}
\def\gev{\mbox{ GeV}}

\def\ln#1{\log{\l( #1 \r)}}

\def\div{\frac{1}{\hat{\epsilon}}}

\def\Im{\mathop{\mbox{Im}}}
\def\Re{\mathop{\mbox{Re}}}

\def\exponential{{\mathrm{e}}}

\usepackage{xcolor}

\begin{document}

\title{Progress in the determination of Mellin moments of the pion LCDA using the HOPE method}
\ShortTitle{Pion LCDA Moments from HOPE Method}

\manuallySeparateAuthors
\author[a,b]{William~Detmold,}
\author*[a]{ Anthony~V.~Grebe,}
\author[c]{ Issaku~Kanamori,}
\author[d,e,f]{ C.-J.~David~Lin,}
\author[g]{ Santanu~Mondal,}
\author*[d,e]{ Robert~J.~Perry}
\author[g]{ and Yong~Zhao}
\author{ for the HOPE collaboration}

\affiliation[a]{Center for Theoretical Physics,\\ Massachusetts Institute of Technology,
Cambridge, MA 02139, USA}
\affiliation[b]{The NSF AI Institute for Artificial Intelligence and Fundamental Interactions}
\affiliation[c]{RIKEN Center for Computational Science,\\ Kobe 650-0047, Japan}
\affiliation[d]{Institute of Physics,\\ National Yang Ming Chiao Tung University, Hsinchu 30010, Taiwan}
\affiliation[e]{Centre for Theoretical and Computational Physics,\\ National Yang Ming Chiao Tung University, Hsinchu 30010, Taiwan}
\affiliation[f]{Centre for High Energy Physics,\\ Chung-Yuan Christian University, Chung-Li, 32032, Taiwan}
\affiliation[g]{Los Alamos National Laboratory,\\ Theoretical Division T-2, Los Alamos, NM 87545, USA}
\affiliation[g]{Physics Division, Argonne National Laboratory, Lemont, IL 60439, USA}

\emailAdd{wdetmold@mit.edu}
\emailAdd{agrebe@mit.edu}
\emailAdd{kanamori-i@riken.jp}
\emailAdd{dlin@nycu.edu.tw}
\emailAdd{santanu.sinp@gmail.com}
\emailAdd{perryrobertjames@gmail.com}

\abstract{
The pion light-cone distribution amplitude (LCDA) is a central non-perturbative object of interest for the calculation of high-energy exclusive processes in quantum chromodynamics. 
In this article, we discuss the calculation of the second and fourth Mellin moment of the pion LCDA using a heavy-quark operator product expansion. The resulting value for the second Mellin moment is $ \expval{ \xi^2 }(\mu = 2~\text{GeV})= 0.210 \pm 0.013\text{ (stat.)} \pm 0.034\text{ (sys.)}$.
This result is compatible with those from previous determinations of this quantity. 
}

\FullConference{%
 The 38th International Symposium on Lattice Field Theory, LATTICE2021
  26th-30th July, 2021
  Zoom/Gather@Massachusetts Institute of Technology
}
\maketitle

\section{Introduction}
\label{sec:intro}
The pion light-cone distribution amplitude (LCDA) plays an important role in parton physics and is central to a description of a range of exclusive processes in high energy quantum chromodynamics~\cite{Radyushkin:1977gp,Lepage:1980fj}. The pion LCDA is denoted $\phi_{\pi}$ and can be defined via the matrix element for the transition between the vacuum and the
(charged\footnote{The isospin limit is used for the calculation  presented in this article.}) pion state,
\beq
\label{eq:pion_DA_def}
\la 0 | \overline{\psi}_d(z) \gamma_{\mu} \gamma_{5} W[z, -z] \psi_u(-z) |
\pi^{+} ({\mathbf{p}}) \ra =
 i  p_{\mu} f_{\pi} \int_{-1}^{1} d \xi \mbox{ }
 {\mathrm{e}}^{-i \xi p\cdot z }\phi_{\pi}(\xi, \mu ) \, ,
\eeq
where
$\mu$ is the renormalization scale and ${\mathcal{W}}[-z,z]$ is a light-like Wilson line connecting $-z$ and $z$.  In the above equation,
$f_{\pi}$ is the decay constant and $p_{\mu}$ is the four-momentum of the pion.  In the light-cone gauge, $\phi_{\pi}(\xi, \mu )$ can
be interpreted as the probability amplitude to convert the pion into a state of a quark and an antiquark carrying momentum fractions $(1+\xi)/2$ and $(1-\xi)/2$, respectively~\cite{Lepage:1980fj}. 

Moments of the LCDA
can be extracted by computing matrix elements of local operators that result from an
operator product expansion (OPE)~\cite{Kronfeld:1984zv, Martinelli:1987si,Del_Debbio_2003,Arthur_2011,Bali:2019dqc}.  In principle, knowledge of the
relevant Mellin
moments enables the construction of PDFs and LCDAs.   More recently, alternative approaches have been proposed to extract information about LCDAs and PDFs using lattice QCD (LQCD)~\cite{Aglietti:1998ur,Liu:1999ak,Detmold:2005gg,Braun:2007wv,Davoudi:2012ya,Ji:2013dva,Chambers:2017dov,Radyushkin:2017cyf,Ma:2017pxb}. The implementations of these approaches have been reviewed in Refs.~\cite{Cichy:2018mum,Ji:2020ect,Constantinou:2020pek,Constantinou:2020hdm,Cichy:2021lih}.

This article reports on recent progress on the study of the Mellin moments of the pion using the heavy-quark operator product expansion (HOPE) method~\cite{Detmold:2005gg}. In this approach, one calculates a hadronic amplitude in LQCD where standard currents are replaced by fictitious heavy-light flavor-changing currents. The numerical matrix element is then related to the moments of the LCDA via an operator product expansion (OPE). It is possible to show that the effect of the heavy quark may be absorbed in the definition of the Wilson coefficients~\cite{Detmold:2021uru}. It is expected that the use of a heavy quark leads to several advantages including the removal of certain higher-twist effects, additional control of the OPE due to the additional hard scale provided by the heavy quark mass, and the simple analytic continuation to Minkowski space~\cite{Detmold:2020lev,Detmold:2021uru}. Since this is the first numerical implementation of this approach, the
quenched approximation and an unphysical quark mass corresponding to a pion mass of approximately $550\mev$ are used.

The structure of this article is as follows: details of the matrix elements which must be calculated are summarized in Sec.~\ref{sec:strategy}, details of the numerical implementation are given in Sec.~\ref{sec:numerical}, analysis of the second moment is provided in Sec.~\ref{sec:analysis}, exploratory analysis of the fourth moment is given in Sec.~\ref{sec:fourth_moment_analysis} and conclusions and outlook are provided in Sec.~\ref{sec:conclusion}.

\section{Strategy and correlation functions}
\label{sec:strategy}
\label{sec:correlators}

Since the light-cone collapses to a point in Euclidean space, the light-like separation $z$ prevents direct computation of the pion LCDA from its definition in Eq.~(\ref{eq:pion_DA_def}).  Instead, one can define the hadronic amplitude
\begin{equation}
\label{eq:hadronic_tensor}
V^{\mu\nu} (q,p) = \int d^{4}z \mbox{ } \exponential^{i q \cdot z} \mbox{ }
 \la 0 | {\mathcal{T}} \{ J_{A}^{\mu} (z/2) J_{A}^{\nu}(-z/2) \} | \pi ({\mathbf{p}})\ra \, ,
\end{equation}
where
\begin{equation}
\label{eq:axial_current}
 J^{\mu}_{A} = \overline{\Psi} \gamma^{\mu}\gamma^{5}\psi + \overline{\psi} \gamma^{\mu}\gamma^{5} \Psi \, ,
\end{equation}
is an axial-vector current that converts between the light quark $\psi$ and 
a valence heavy quark $\Psi$ with mass $m_{\Psi}$.

The antisymmetrized hadronic amplitude $V^{[\mu\nu]} = (V^{\mu\nu}-V^{\nu\mu})/2$ can be expanded in terms of an operator product expansion,
\begin{equation}
    V^{[\mu\nu]} (q,p) = -\frac{2i\epsilon^{\mu\nu\rho\sigma} q_\rho p_\sigma}{\tilde Q^2} f_\pi \sum_{n=0,\text{even}}^\infty C_W^{(n)} (\tilde Q^2, \mu, m_\Psi)\langle \xi^n \rangle \left( \frac{\tilde \omega}{2} \right)^n \, ,
    \label{eq:OPE}
\end{equation}
where $\tilde \omega = 2 p\cdot q/\tilde Q^2$, $\tilde Q^2 = -q^2 - m_\Psi^2$, and the $C_W^{(n)}$ are Wilson coefficients that account for short-distance perturbative corrections.  $\langle \xi^n \rangle$ are Mellin moments of the LCDA, defined as
\begin{equation}
    \langle \xi^n \rangle = \int_{-1}^1 d\xi \, \xi^n \phi_\pi(\xi, \mu) \, .
\end{equation}
The zeroth moment $\langle \xi^0 \rangle$ is 1, and all odd moments vanish by the isospin symmetry assumed here.  Given full knowledge of the moments, it is possible to reconstruct the entire functional dependence of the LCDA, and even partial knowledge is useful for constraining the LCDA.

The hadronic amplitude can also be written in terms of quantities calculable on the lattice.
Defining
\begin{align}
  R^{\mu\nu}(\tau; \mathbf{p}, \mathbf{q})
 &= \int d^3 \mathbf{z} \,e^{i\mathbf{q}\cdot \mathbf{z}} \langle 0 |\mathcal{T}[ J_A^\mu(\tau/2,\mathbf{z}/2) J_A^\nu(-\tau/2,-\mathbf{z}/2)] | \pi(\mathbf{p}) \rangle
 \nonumber \\
 &= \langle 0 | J_A^\mu(\tau/2; (\mathbf{p}+\mathbf{q})/2) J_A^\nu(-\tau/2; (\mathbf{p}-\mathbf{q})/2) | \pi(\mathbf{p}) \rangle \, ,
  \label{ratio}
\end{align}
and performing a Fourier transform in time gives the hadronic amplitude
\begin{equation}
  V^{\mu\nu}(q,p) = \int d\tau \, e^{iq_4 \tau} R^{\mu\nu}(\tau; \mathbf{p}, \mathbf{q}) \, .
  \label{fourier-transform-temporal}
\end{equation}

\begin{figure}
  \centering
  \includegraphics[scale=0.7]{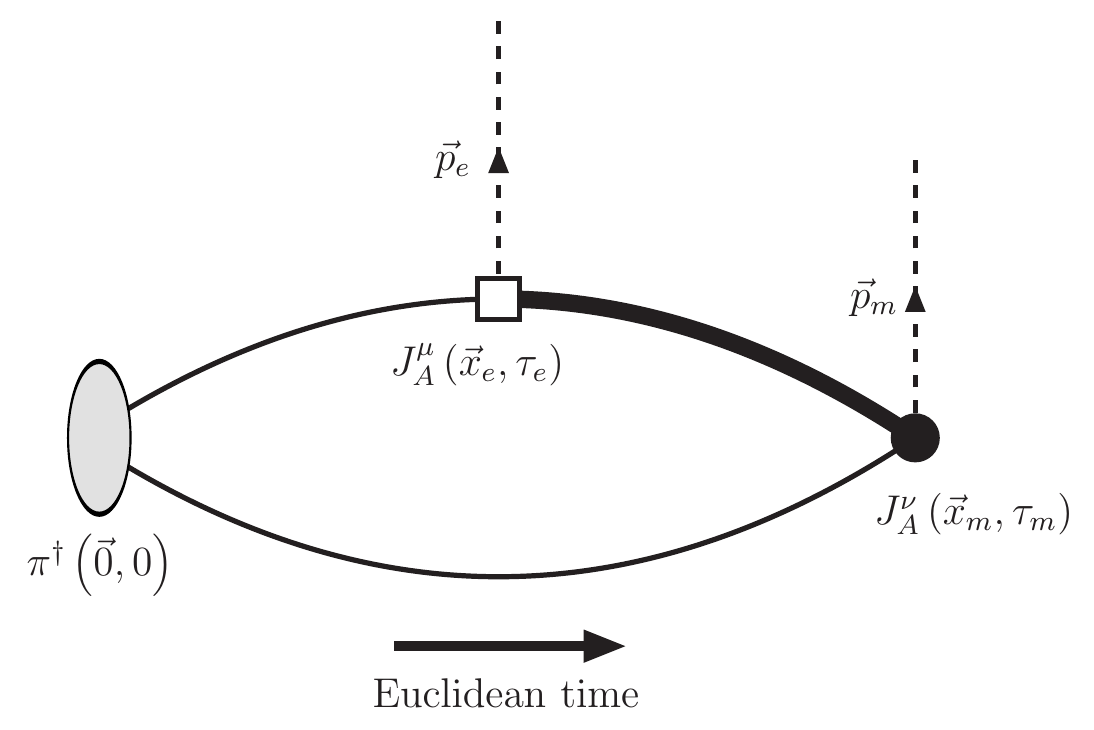}
  \caption{A diagram of the three-point correlator used in this calculation with current insertions at times $\tau_e$ and $\tau_m$.}
  \label{3-pt-figure}
\end{figure}

To extract $R^{\mu\nu}$ on the lattice, one starts by computing two-point and three-point correlators
\begin{align}
  C_2 (\tau, \mathbf{p})
  &= \int d^3 \mathbf{x} \, e^{i\mathbf{p}\cdot \mathbf{x}}  \langle 0|\mathcal{O}_\pi(\tau,\mathbf{x}) \mathcal{O}^\dagger_\pi (0, \mathbf{0}) |0 \rangle \nonumber \\
  &= \langle 0 | \mathcal{O}_\pi(\tau, \mathbf{p})
    \mathcal{O}^\dagger_\pi(0, \mathbf{p}) | 0 \rangle \, ,
  \label{2pt-corr}
\end{align}
and
\begin{align}
  C^{\mu\nu}_3 (\tau_e, \tau_m; \mathbf{p}_e, \mathbf{p}_m)
  &= \int d^3x_e \, d^3x_m\, e^{i\mathbf{p}_e\cdot \mathbf{x}_e}e^{i\mathbf{p}_m\cdot \mathbf{x}_m}\langle 0 | \mathcal{T}\left[ J_A^\mu(\tau_e, \mathbf{x}_e) J_A^\nu(\tau_m, \mathbf{x}_m) \mathcal{O}^\dagger_\pi(\mathbf{0}) \right] | 0 \rangle
  \nonumber \\
  &= \langle 0 | J_A^\mu(\tau_e, \mathbf{p}_e) J_A^\nu(\tau_m, \mathbf{p}_m) \mathcal{O}^\dagger_\pi(0, \mathbf{p}) | 0 \rangle  \, .
  \label{3pt-corr}
\end{align}
The three-point correlator is shown diagrammatically in Fig.~\ref{3-pt-figure}.

At large source-sink separations, one can extract the pion energy $E_\pi(\mathbf{p})$ and overlap factor $Z_\pi(\mathbf{p}) = \langle 0 | \mathcal{O}_\pi | \pi(\mathbf{p}) \rangle$ from the two-point function and then construct the ratio
\begin{equation}\label{eq:r12}
  R^{\mu\nu}(\tau; \mathbf{p}, \mathbf{q}) = \frac{2E_\pi(\mathbf{p}) C_3^{\mu\nu}(\tau_e, \tau_m; \mathbf{p}_e, \mathbf{p}_m)}{Z_\pi(\mathbf{p}) e^{-E_\pi(\mathbf{p})(\tau_e + \tau_m)/2}} \, .
\end{equation}
From this and Eqs.~(\ref{eq:OPE}) and (\ref{fourier-transform-temporal}), one can in principle extract any number of Mellin moments on the lattice.

\section{Details of numerical implementation}
\label{sec:numerical}
Order-$a$ improvement of the quark propagators used in constructing $C_2$ and $C_3^{\mu\nu}$ requires inclusion of a clover term in the Wilson action~\cite{Symanzik:1983dc,Symanzik:1983gh,Luscher:1996sc} with non-perturbatively tuned clover coefficient, here taken from Ref.~\cite{Luscher_1997}.  The $\mathcal{O}(a)$ improvement to the axial vector operators used here is a multiplicative factor that cancels in determinations of the second moment~\cite{Detmold:2021qln}.

Higher-twist corrections to the operator product expansion in Eq.~(\ref{eq:OPE}) are proportional to $\Lambda_\text{QCD} / \tilde{Q}$ or $m_\pi / \tilde{Q}$.  
In other approaches, these are suppressed by large values of $\mathbf q^2$ (or small separations in position space)~\cite{Bali_2018}, but this work makes use of the heavy quark mass to make $\tilde Q$ large.  Taking $m_\Psi \gtrsim q_4 \gg |\mathbf{q}|$ implies $\tilde Q \sim m_\Psi$, so suppressing higher-twist effects that scale as $\Lambda_\text{QCD}/m_\Psi$ or $m_\pi/m_\Psi$ requires $\Lambda_\text{QCD}, m_\pi \ll m_\Psi$.  Since one must also control lattice artifacts proportional to a power of either $a \Lambda_\text{QCD}$ or $a m_\Psi$, one must ensure that
\beq
\label{eq:scale_hierarchy}
\Lambda_\text{QCD}, m_\pi \ll m_\Psi \lesssim a^{-1} \, .
\eeq

\begin{figure}[h]
  \centering
  \includegraphics[width=0.49\textwidth]{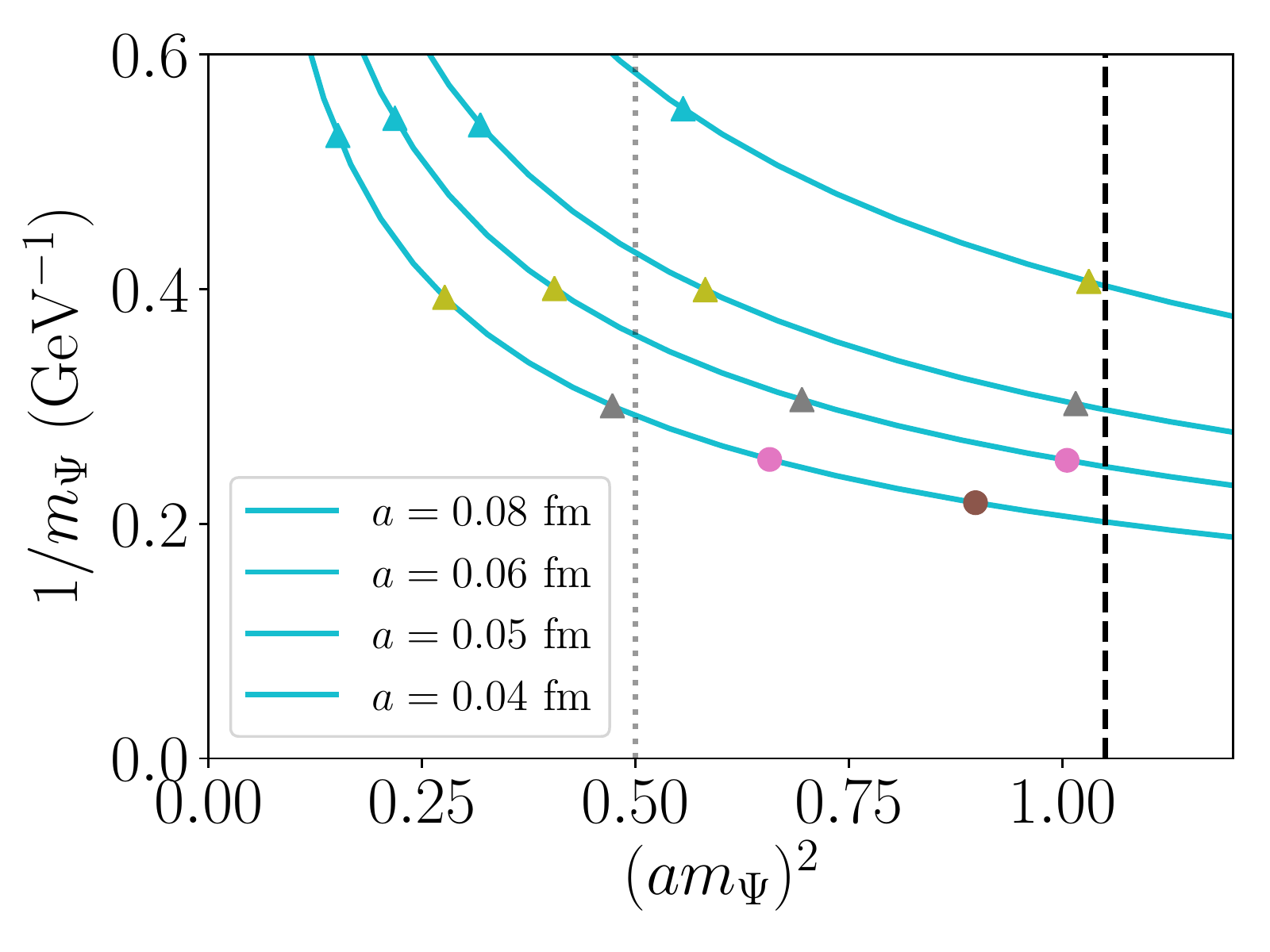}
  \caption{The four lattice spacings used here are represented by the curves, with the finest spacing closest to the origin.  Along each curve, as the heavy quark mass $m_\Psi$ is varied, there is a trade-off between discretization effects that scale as $(am_\Psi)^2$ and higher-twist effects proportional to $1/m_\Psi$.  The masses used in this study are represented by the coloured points, which tuned to be approximately constant with respect to lattice spacing.  They were chosen to be large enough to suppress higher-twist effects (about 1.8 GeV and heavier) and also small enough to control discretization effects ($am_\Psi \leq 1.05$, as shown by the dashed black line).  For analysis of systematic uncertainties from lattice artifacts, the more conservative threshold of $am_\Psi < 0.7$, shown by the dotted gray line, was used.}
  \label{fig:masses-used}
\end{figure}

The heavy quark mass $m_\Psi$ is taken to be at least 1.8 GeV to be significantly larger than both $\Lambda_\text{QCD} \sim 250$ MeV and $m_\pi \approx 550$ MeV.  Even heavier masses, as large as 4.5 GeV, are also used in order to fit away residual higher-twist effects.  Including such masses without uncontrolled lattice artifacts requires lattice spacings as small as 0.04 fm.  The full range of lattice spacings and heavy quark masses used in this work is shown in Fig.~\ref{fig:masses-used}.

At fine lattice spacings, the dominant cost for calculations with dynamical quarks is typically ensemble generation, which is plagued by the problem of critical slowing down.  To circumvent this problem, this preliminary calculation was performed in the quenched approximation with configurations generated via the multiscale approach of Ref.~\cite{Detmold:2018zgk}.  The physical volume was tuned to 1.92 fm, as determined via Wilson flow scale setting \cite{Detmold:2018zgk,asakawa2015determination}.

The pion mass was tuned to the unphysical value of $m_\pi \approx 550$ MeV across ensembles.  This suppresses finite-volume effects, since $m_\pi L \approx 5.3$.  The heavy quark masses were tuned to give the heavy-heavy pseudoscalar meson a constant mass on all four lattice spacings.  The resulting bare masses and lattice spacings are detailed in Table \ref{lattices-used}.  All propagators and contractions were computed using \textsc{Chroma} with the \textsc{QPhiX} inverters \cite{chroma, qphix}.

\begin{table}
\centering
\begin{adjustbox}{width=\textwidth,center=\textwidth}
\begin{tabular}{ c c c c c c c c c } \hline \hline
    $(L/a)^3 \times T/a$ & $\beta$ & $a$ (fm) & $\kappa_\text{light}$ & $\kappa_\text{heavy}$ & $c_\text{sw}$ & Configurations Used & Sources/Config & Total Sources Used \\ \hline
    $24^3 \times 48$ & ~6.10050~ & 0.0813 & ~0.134900~ & $\begin{matrix} 0.1200 \\ 0.1100  \end{matrix}$ & ~1.6842~ & 650 & 12 & 7800 \\ \hline
    $32^3 \times 64$ & 6.30168 & 0.0600 & 0.135154 & $\begin{matrix} 0.1250 \\ 0.1184 \\ 0.1095 \end{matrix}$ & 1.5792 & 450 & 10 & 4500 \\ \hline
    $40^3 \times 80$ & 6.43306 & 0.0502 & 0.135145 & $\begin{matrix} 0.1270 \\ 0.1217 \\ 0.1150 \\ 0.1088 \end{matrix}$ & 1.5292 & 250 & 6 & 1500 \\ \hline
    $48^3 \times 96$ & 6.59773 & 0.0407 & 0.135027 & $\begin{matrix} 0.1285 \\ 0.1244 \\ 0.1192 \\ 0.1150 \\ 0.1100 \end{matrix}$ & 1.4797 & 341 & 10 & 3410 \\ \hline
  \end{tabular}
\end{adjustbox}
  \caption{Ensembles and quark masses studied in this work. }
  \label{lattices-used}
\end{table}

\subsection{Techniques for Signal Improvement}

The signal-to-noise ratio is largest for pion momenta that are as small as possible, so $\mathbf{\hat p} \equiv \frac{\mathbf{p}}{2\pi/L}$ was chosen to correspond to one unit of momentum (about 640 MeV) for the second-moment calculation.  As a consequence, the prefactor of $\langle \xi^2 \rangle$ in Eq.~(\ref{eq:OPE}) is about $10^{-2}$ at the lightest $m_\Psi$ and even smaller for heavier masses.
This small contribution must be isolated from the $O(1)$ zeroth-moment contribution.  Fixing the axial current indices to be $\mu=1, \nu=2$,
the overall prefactor of the OPE in Eq.~(\ref{eq:OPE}) can be written as
\beq
\label{eq:prefactor_our_kinematics}
i\varepsilon_{\mu\nu\rho\sigma}q^\rho p^\sigma = i\left(q^0 p^3 - p^0 q^3\right) = - q^4 p^3 - i E_\pi q^3 \, .
\eeq
Choosing $p^3 = 0$, the overall prefactor times $\langle \xi^0\rangle = 1$ is purely imaginary.  By also choosing $\mathbf{p}\cdot \mathbf{q} \neq 0$, $p\cdot q = iE_\pi q^4 - \mathbf{p}\cdot \mathbf{q}$ to be complex, the second moment contribution is given a nonzero real part.  As a result, the real part of the hadronic amplitude is dominated by the contribution of $\langle \xi^2 \rangle$.

Since one can show that $R^{\mu\nu}(\tau)$ is also pure imaginary \cite{Detmold:2021qln}, 
properties of the Fourier transform imply that the imaginary and real parts of $V^{\mu\nu}$ correspond to symmetric and antisymmetric combinations of $R^{\mu\nu}(\pm \tau)$:
\begin{align}
  \text{Re}[V^{\mu\nu}(\mathbf{p}, q)] &= \int_0^\infty d\tau \, \left[ R^{\mu\nu}(\tau; \mathbf{p}, \mathbf{q}) - R^{\mu\nu}(-\tau; \mathbf{p}, \mathbf{q}) \right] \sin (q_4 \tau) \, ,
  \label{antisymmetry-R} \\ 
  \text{Im}[V^{\mu\nu}(\mathbf{p}, q)] &= \int_0^\infty d\tau \, \left[ R^{\mu\nu}(\tau; \mathbf{p}, \mathbf{q}) + R^{\mu\nu}(-\tau; \mathbf{p}, \mathbf{q}) \right] \cos (q_4 \tau) \, .
  \label{symmetry-R}
\end{align}
The second moment contribution and therefore the real part of $V^{\mu\nu}$ are suppressed by about two orders of magnitude relative to the imaginary part.  Therefore, the difference in Eq.~(\ref{antisymmetry-R}) involves large cancellations between the two terms in the integral.  Statistical fluctuations in the correlated difference can be much larger than fluctuations in the individual terms if the terms are highly correlated.  One can increase these correlations by computing $C_3^{\mu\nu}(\tau < 0)$ via the identity
\begin{equation}
  C_3^{\mu\nu}(\tau_e, \tau_m; \mathbf{p}_e, \mathbf{p}_m)^* = C_3^{\nu\mu}(\tau_m, \tau_e; -\mathbf{p}_m, -\mathbf{p}_e) \, ,
\label{c3-identity}
\end{equation}
which can be proved using $\gamma_5$-hermiticity of the quark propagator.
Applying this identity to Eq.~(\ref{antisymmetry-R}) and (\ref{symmetry-R}) yields
\begin{align}
  \text{Re}[V^{\mu\nu}(\mathbf{p}, q)] &= \int_0^\infty d\tau \, \left[ R^{\mu\nu}(\tau; \mathbf{p}, \mathbf{q}) + R^{\mu\nu}(\tau; -\mathbf{p}, \mathbf{q}) \right] \sin (q_4 \tau) \, , \label{antisymmetry-R-improved} \\
  \text{Im}[V^{\mu\nu}(\mathbf{p}, q)] &= \int_0^\infty d\tau \, \left[ R^{\mu\nu}(\tau; \mathbf{p}, \mathbf{q}) - R^{\mu\nu}(\tau; -\mathbf{p}, \mathbf{q}) \right] \cos (q_4 \tau) \, . \label{symmetry-R-improved}
\end{align}

Using this identity gives access to both $\tau > 0$ and $\tau < 0$ from a single set of $\tau_e, \tau_m$ values, which will enhance the correlations between spatially localized statistical fluctuations and therefore decrease the statistical uncertainty.

Excited state contamination was observed to fall to the $\sim 1\%$ level by $\tau_e \approx 0.7$ fm.  Therefore, the smaller of the two current insertion times will be fixed to 0.7 fm in this analysis.

\section{Second Moment}
\label{sec:analysis}
As discussed above, both lattice artifacts and higher-twist effects contaminate the signal, and different quark masses explore different trade-offs between these.  Determining $\langle \xi^2 \rangle$ from the 2- and 3-point correlators and removing these sources of contamination from the signal is nontrivial, so two separate analyses, referred to below as \textit{time-momentum analysis} and the \textit{momentum-space analysis}, have been performed to ensure that systematics are under control.  These methods can be summarized as follows:
\begin{enumerate}
  \item Time-momentum analysis
  \begin{enumerate}[label=(\roman*)]
    \item Perform a continuous inverse Fourier transform of the OPE of  $\text{Im}[V(\mathbf{p},q)]$, as described in Eq.~(\ref{symmetry-R-improved}), to convert it to the time-momentum representation.  Compare this to the ratio of lattice correlators to fit the pion decay constant $f_\pi$ and heavy quark mass $m_\Psi$.
    \item Perform a continuous inverse Fourier transform of $\text{Re}[V(\mathbf{p},q)]$, shown in Eq.~(\ref{antisymmetry-R-improved}), and use $f_\pi, m_\Psi$ as inputs to fit $\langle \xi^2\rangle$ at each heavy quark mass and lattice spacing.
    \item Remove both discretization and higher-twist artifacts via a global fit of $\langle \xi^2 \rangle(a, m_\Psi)$ computed from all lattice data.
  \end{enumerate}
  \item Momentum-space analysis
  \begin{enumerate}[label=(\roman*)]
    \item Construct the ratio of correlators $R^{\mu\nu}(\tau)$ from lattice data and perform a discrete Fourier transform to convert to the momentum-space hadronic amplitude.
    \item At each heavy quark mass, perform a continuum extrapolation of the hadronic amplitude.
    \item Fit the continuum-limit hadronic amplitude to the HOPE formula given in Eq.~(\ref{eq:OPE}) augmented by a model of higher-twist effects to extract $f_\pi$, $m_\Psi$ and $\expval{\xi^2}$.
  \end{enumerate}
\end{enumerate}
As demonstrated in Ref.~[28], the two analysis methods yield compatible results, but the time-momentum procedure results in a slightly smaller systematic error.  Therefore, it is presented in detail below.  A comprehensive description of the momentum-space analysis can be found in Ref.~[28].

\subsection{Time-Momentum Analysis}
\label{time-momentum}

\begin{figure}
\centering
\subfloat[][]{
    \includegraphics[width=0.45\textwidth]{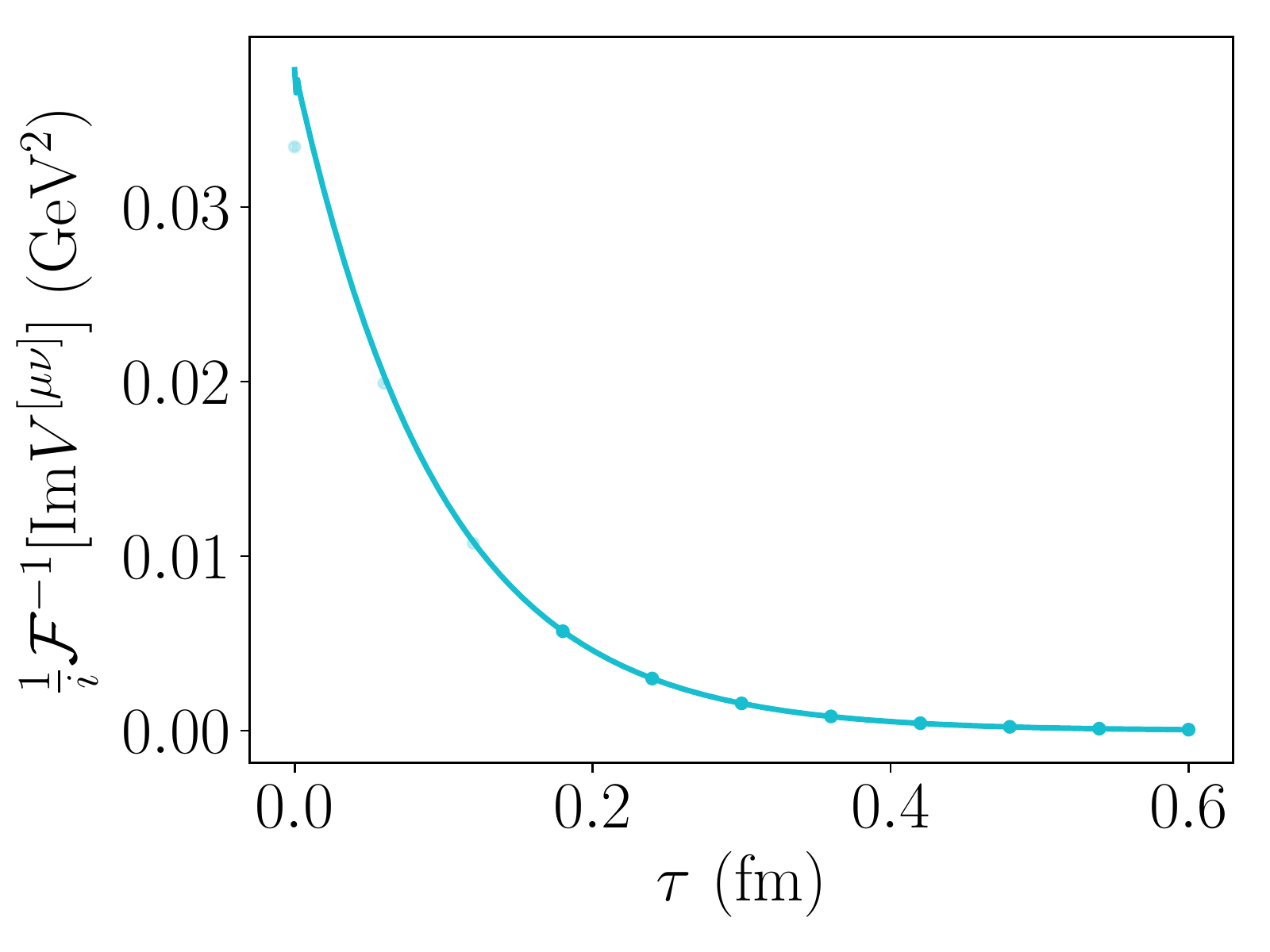}
}
\subfloat[][]{
    \includegraphics[width=0.45\textwidth]{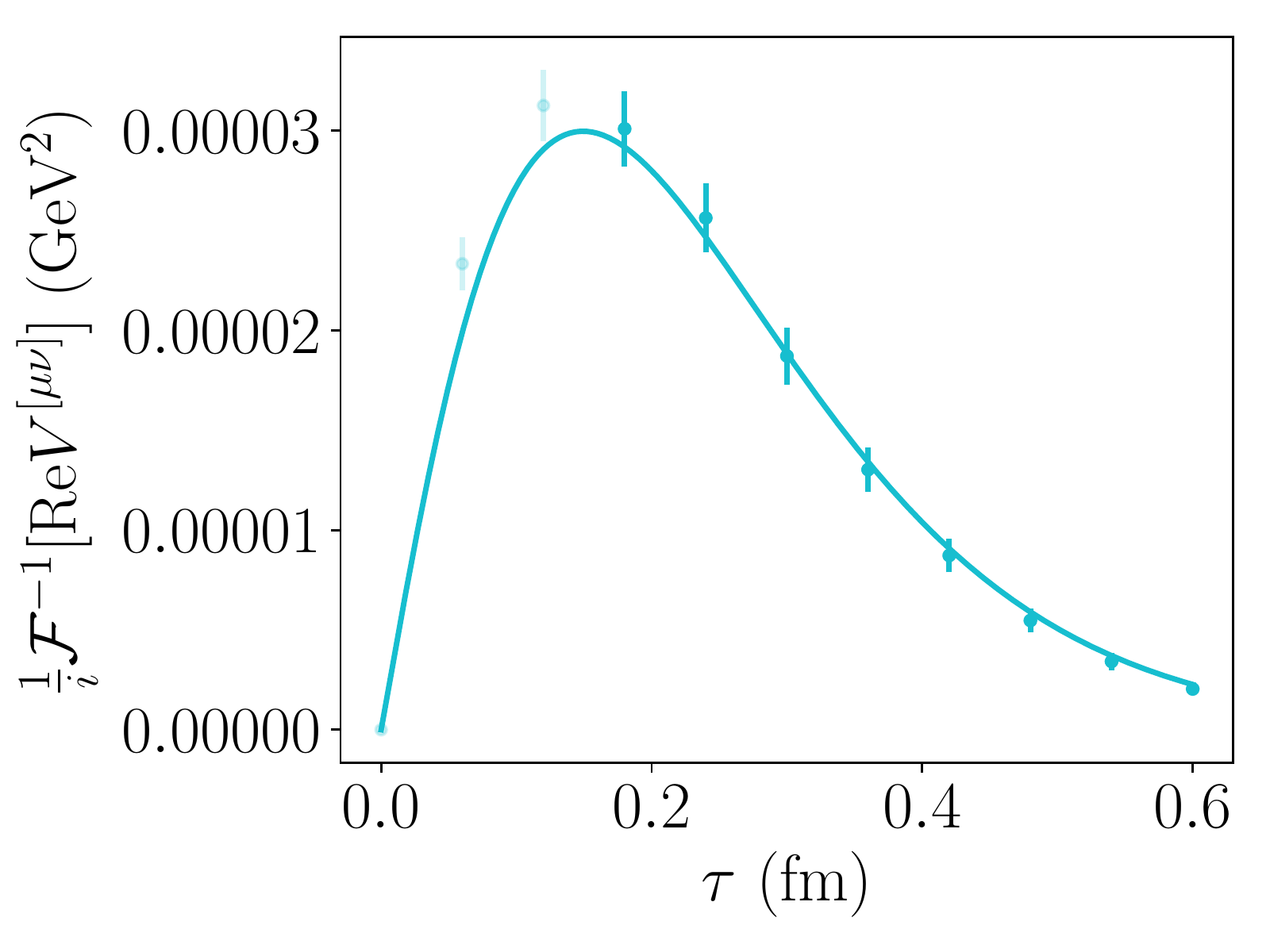}
}
\caption{Inverse Fourier transforms of the imaginary and real parts of the antisymmetrized hadronic tensor $V^{[\mu\nu]}$.  The light shaded points are not used in the fit due to contamination from discretization effects scaling as $a/\tau$.  (a) $f_\pi$ and $m_\Psi$ can be extracted from the symmetric part of $R^{[\mu\nu]}(\tau)$ (equal to $\mathcal{F}^{-1}[\text{Im}(V^{[\mu\nu]})]$), which is dominated by the zeroth moment contribution.  (b) Using these as inputs, the second moment $\langle \xi^2 \rangle$ can be fit to $\mathcal{F}^{-1}[\text{Re}(V^{[\mu\nu]})]$.}
  \label{fig:single-mass}
\end{figure}

The ratio $R^{[\mu\nu]}\left( \tau; \mathbf{p}, \mathbf{q} \right)$ was
constructed from lattice correlators for $3a \leq \tau \lesssim 0.6$ fm, where the lower bound suppresses contributions from discretization effects scaling as $a/\tau$ and the upper bound removes data with larger statistical uncertainties and higher-twist contamination.
$R^{[\mu\nu]}$ is then fit to the inverse Fourier transform of $V^{[\mu\nu]}$ at a given lattice spacing and $m_\Psi$, as shown in Fig.~\ref{fig:single-mass}.  Since the imaginary part of the second moment contribution is negligible, the fitting procedure splits into two steps: extracting $f_\pi$ and $m_\Psi$ from the imaginary part of $V^{[\mu\nu]}$ and then using $f_\pi$ and $m_\Psi$ as inputs for fitting $\langle \xi^2 \rangle$ from $\text{Re}[V^{[\mu\nu]}]$.

\begin{figure}
  \centering
  \includegraphics[width=0.49\textwidth]{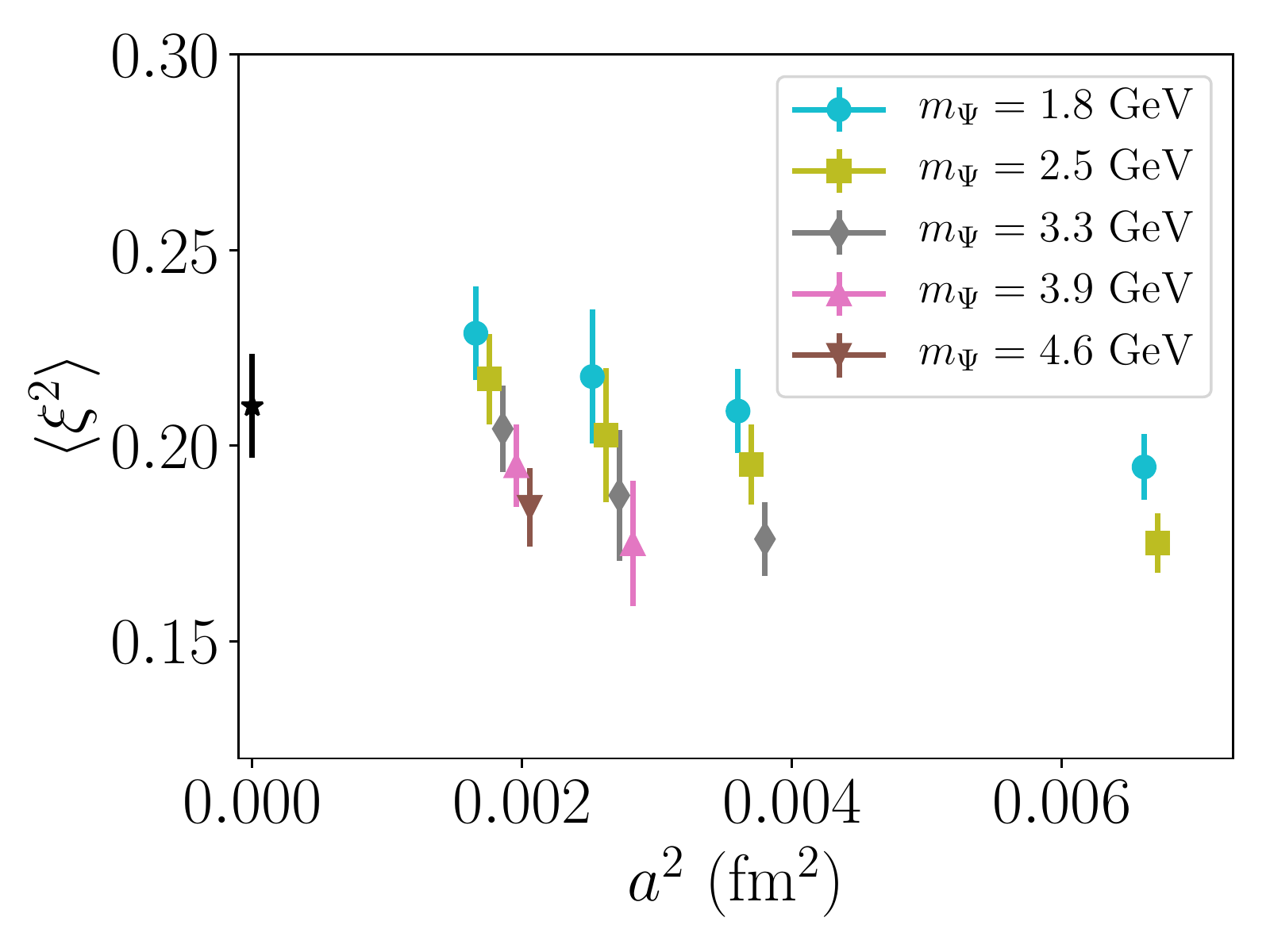}
\caption{Extracted values of $\langle \xi^2 \rangle(a,m_\Psi)$ at finite lattice spacing and heavy quark mass shows significant variation due to discretization and higher-twist corrections.  Only when these effects have been removed by fitting to Eq.~(\ref{eqn:global-fit}) does one obtain the continuum, twist-2 value shown the black star at $a^2=0$.}
  \label{fig:global-fit}
\end{figure}

The second moment is extracted from comparison of lattice data with both discretization artifacts and higher-twist corrections to an OPE expressed without them in Eq.~(\ref{eq:OPE}), so the extracted value $\langle \xi^2 \rangle (a, m_\Psi)$ is contaminated with these artifacts at any non-zero lattice spacing.
For the $O(a)$-improved action, lattice artifacts start at $O(a^2)$ times powers of the heavy quark mass $m_\Psi$.
At pion masses comparable to $\Lambda_\text{QCD}$ and with $q_4 \sim m_\Psi \sim \tilde Q$, higher-twist effects scale as powers of $\Lambda_\text{QCD}/\tilde{Q} \sim \Lambda_\text{QCD}/m_\Psi$.  Together, these artifacts imply that $\langle \xi^2 \rangle (a, m_\Psi)$ can be fit to 
\begin{equation}
  \langle \xi^2 \rangle(a,m_\Psi) = \langle \xi^2 \rangle + \frac{A}{m_\Psi} + B a^2 + C a^2 m_\Psi + D a^2 m_\Psi^2 \, ,
  \label{eqn:global-fit}
\end{equation}
which determines $\langle \xi^2 \rangle (\mu = 2\text{ GeV}) = 0.210 \pm 0.013$ (stat.) is the continuum, twist-2 value of interest and $A$, $B$, $C$, and $D$ are nuisance parameters.  This extraction is shown in Figure \ref{fig:global-fit}.

\subsection{Estimation of Systematic Uncertainties for Time-Momentum Analysis}

Even after performing the continuum extrapolation, there are other systematic errors that one must account for.
Excited state contamination in the 3-point function is a $\sim 1\%$ effect, and finite-volume effects are negligible here.
The unphysical pion mass of $m_\pi \approx 550$ MeV likely induces a $\sim 5\%$ error in $\langle \xi^2 \rangle$ based on previous studies at various pion masses \cite{Braun_2015}.
Varying the analysis procedure allows one to estimate other systematic uncertainties.
\begin{itemize}
  \item The global fit described in Eq.~(\ref{eqn:global-fit}) was based on a fairly permissive cut of $am_\Psi < 1.05$ for the heavy quark mass, so there could be remaining $O(a^3)$ lattice artifacts.  Taking a more conservative threshold of $am_\Psi < 0.7$ to suppress these higher powers shifts the fitted $\langle \xi^2 \rangle$ by 0.016, which is taken as an estimate of the systematic uncertainty from the continuum extrapolation.
  \item The global fit contains a twist-3 term proportional to $\Lambda_\text{QCD}/m_\Psi$.  There would also be even higher-twist terms in a full OPE, so one could add a $(\Lambda_\text{QCD}/m_\Psi)^2$ term to the global fit in Eq.~(\ref{eqn:global-fit}).  The higher-twist systematic uncertainty is taken to be the shift of 0.025 in $\langle \xi^2\rangle$ when this extra term is included.
  \item At $\tau = \tau_m - \tau_e \leq 2a$, lattice artifacts scaling as $a/\tau$ are uncontrolled and contaminate the signal.  One can take a more conservative cut and exclude $\tau/a = 3$ from the fits, which gives a small systematic uncertainty from the difference in $\langle \xi^2\rangle$ of 0.002.
  \item The Wilson coefficients $C_W$ used in this analysis are calculated to 1-loop order in perturbation theory.  To estimate the effects of higher-loop corrections, one can use a larger renormalization scale of $\mu = 4$ GeV and then evolve the fitted $\langle \xi^2\rangle$ back to $\mu = 2$ GeV, giving a systematic uncertainty of 0.008 from the change in central value.
\end{itemize}

The systematic uncertainties detailed above are summarized in Table~\ref{tab:error-budget} and can be combined into a final result of $\langle \xi^2 \rangle (\mu = 2\text{ GeV}) = 0.210 \pm 0.013\text{ (stat.)} \pm 0.034\text{ (sys.)}= 0.210 \pm 0.036$ (total, exc. quenching). 
The largest contributions are from the continuum and higher-twist extrapolations, and control over both could be improved with finer lattice spacings and therefore heavier masses.
The error from quenching is uncontrollable without repeating these calculations on dynamical ensembles, although in lattice calculations of most spectral quantities, it is typically a 10--20\% effect.

\begin{table}
  \centering
    \begin{tabular}{l c} \hline\hline
    Source of error & Size \\ \hline
    Statistical & 0.013 \\ 
    Discretization & 0.016 \\
    Higher-twist & 0.025 \\
    Excited states & 0.002 \\
    Unphysical $m_\pi$ & 0.014 \\
    Fit range & 0.002 \\
    Higher loops & 0.008 \\\hline
    \textbf{Total (exc.~quenching)} & \textbf{0.036} \\ \hline
  \end{tabular}
  \caption{The various contributions to the uncertainty in this demonstration of the HOPE method to determine the second moment $\langle \xi^2 \rangle$ of the pion LCDA from lattice QCD.}
  \label{tab:error-budget}
\end{table}

\section{Fourth Moment}
\label{sec:fourth_moment_analysis}
The success of the second moment analysis suggests that extractions of higher moments of the pion LCDA may also be possible using this approach. In this section, the first steps towards a systematic study of the fourth Mellin moment are discussed. As explained elsewhere~\cite{Detmold:2020lev,Detmold:2021uru}, higher moments in the HOPE formula are kinematically suppressed by increasing powers of $\tilde{\omega}=2p\cdot q/\tilde{Q}^2$. Thus, like other approaches~\cite{Ji:2013dva,Radyushkin:2017cyf,Ma:2017pxb,Chambers:2017dov}, the statistical sensitivity to higher moments of the LCDA requires calculations at higher hadronic momenta. 

\begin{table}
\centering
\begin{adjustbox}{width=1.\textwidth,center=\textwidth}
\begin{tabular}{ c c c c c c c c } \hline \hline
    $(L/a)^3 \times T/a$ & $\beta$ & $a$ (fm) & $\kappa_\text{light}$ & $\kappa_\text{heavy}$ & $c_\text{sw}$ & No. Configurations &  Total Measurements 
    \\ \hline
    $24^3 \times 48$ & ~6.10050~ & 0.0813 & ~0.134900~ & $\begin{matrix} 0.1200 \\ 0.097 \end{matrix}$ & ~1.6842~ & 3150 & 3150 
    \\ \hline
  \end{tabular}
\end{adjustbox}
  \caption{Details of the ensembles used in the analysis of the fourth Mellin moment.}
  \label{tab:mom_lattices_used}
\end{table}

\subsection{Momentum Smearing}

\begin{figure}
    \centering
    \includegraphics[width=0.49\textwidth]{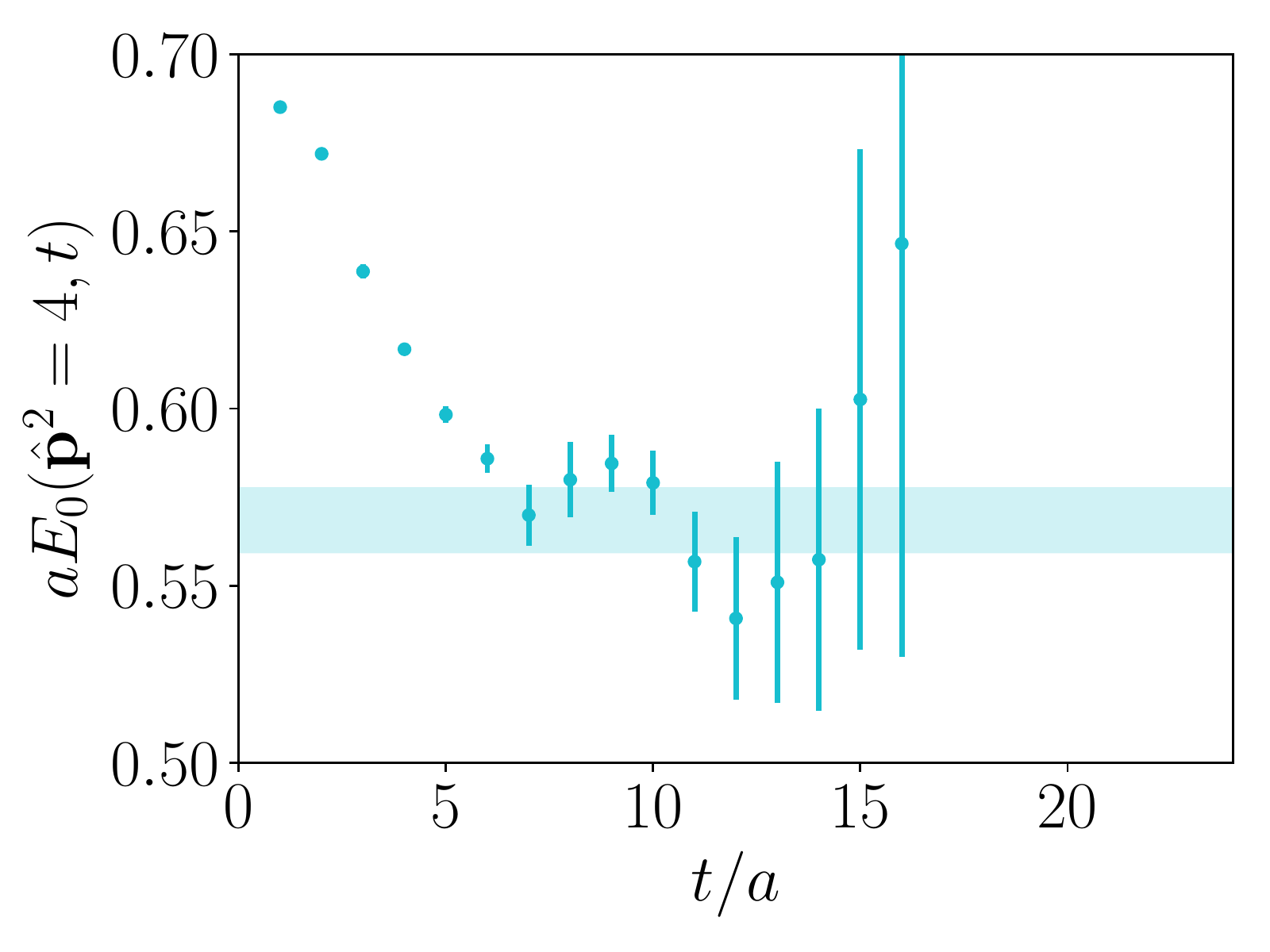}
    \caption{Effective mass calculation for pseudoscalar channel with momentum smeared interpolating operators at a hadronic boost of $|\mathbf{p}|\sim1.3\gev$. The shaded band is the 1-$\sigma$ curve for the fitted ground state energy.}
    \label{fig:high_mom_2pt}
\end{figure}

Calculating hadronic matrix elements at large momenta is well known to be a challenging problem in LQCD due to the exponentially decreasing signal-to-noise problem, and also due to enhanced excited state contamination. In order to reduce this excited state contamination, momentum smearing~\cite{Bali:2016lva} is used to reduce the coupling of the interpolating operators to excited states. 
In the free-field case, momentum smearing amounts to the convolution of the quark source with a Fourier phase factor $e^{i\mathbf{k}\cdot\mathbf{x}}$, and thus results in the re-centering of the three-momentum distribution of the quark propagator around $\mathbf{k}$. Following the notation of Ref.~\cite{Bali:2016lva}, it was found that the best reduction of excited states was obtained with $\zeta=0.8$ and $\mathbf{k}=\zeta\mathbf{p}$, where $\mathbf{p}$ is the three-momentum of the pion. An example of the resulting two-point correlator is shown in Fig.~\ref{fig:high_mom_2pt}.

\subsection{Excited state contamination}

\begin{figure}
    \centering
    \includegraphics[width=0.49\textwidth]{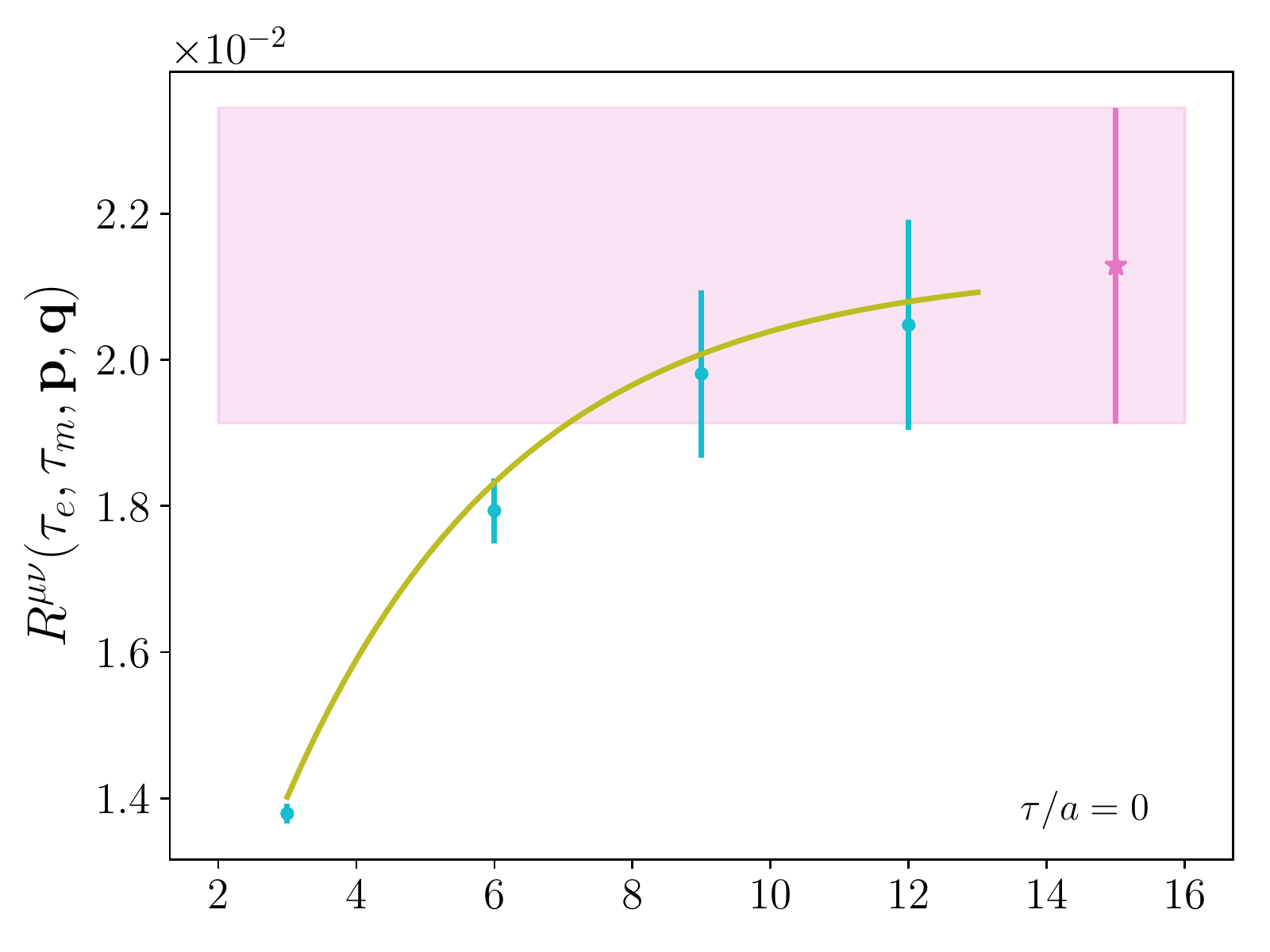}
    \includegraphics[width=0.49\textwidth]{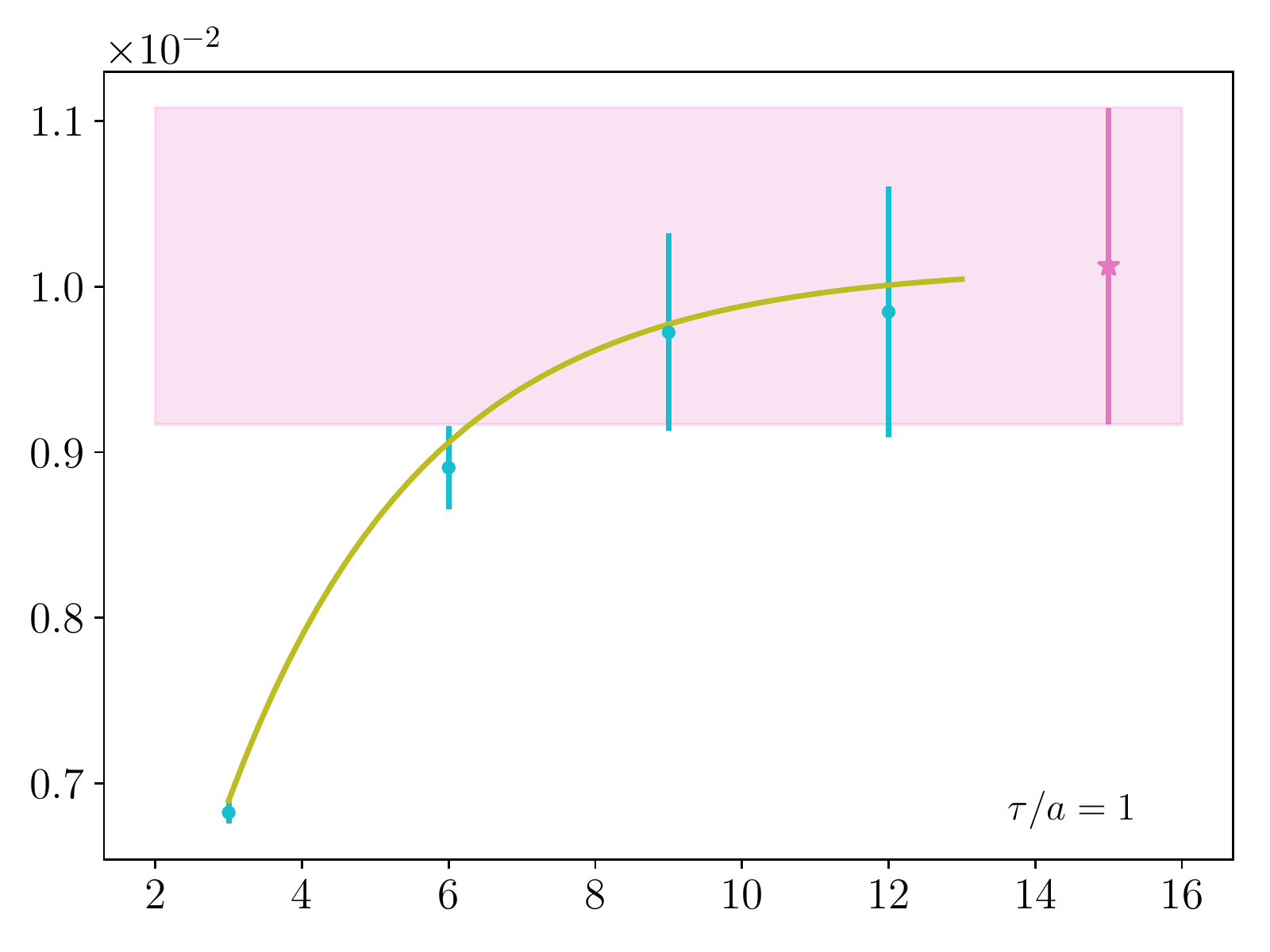}
    
    \includegraphics[width=0.49\textwidth]{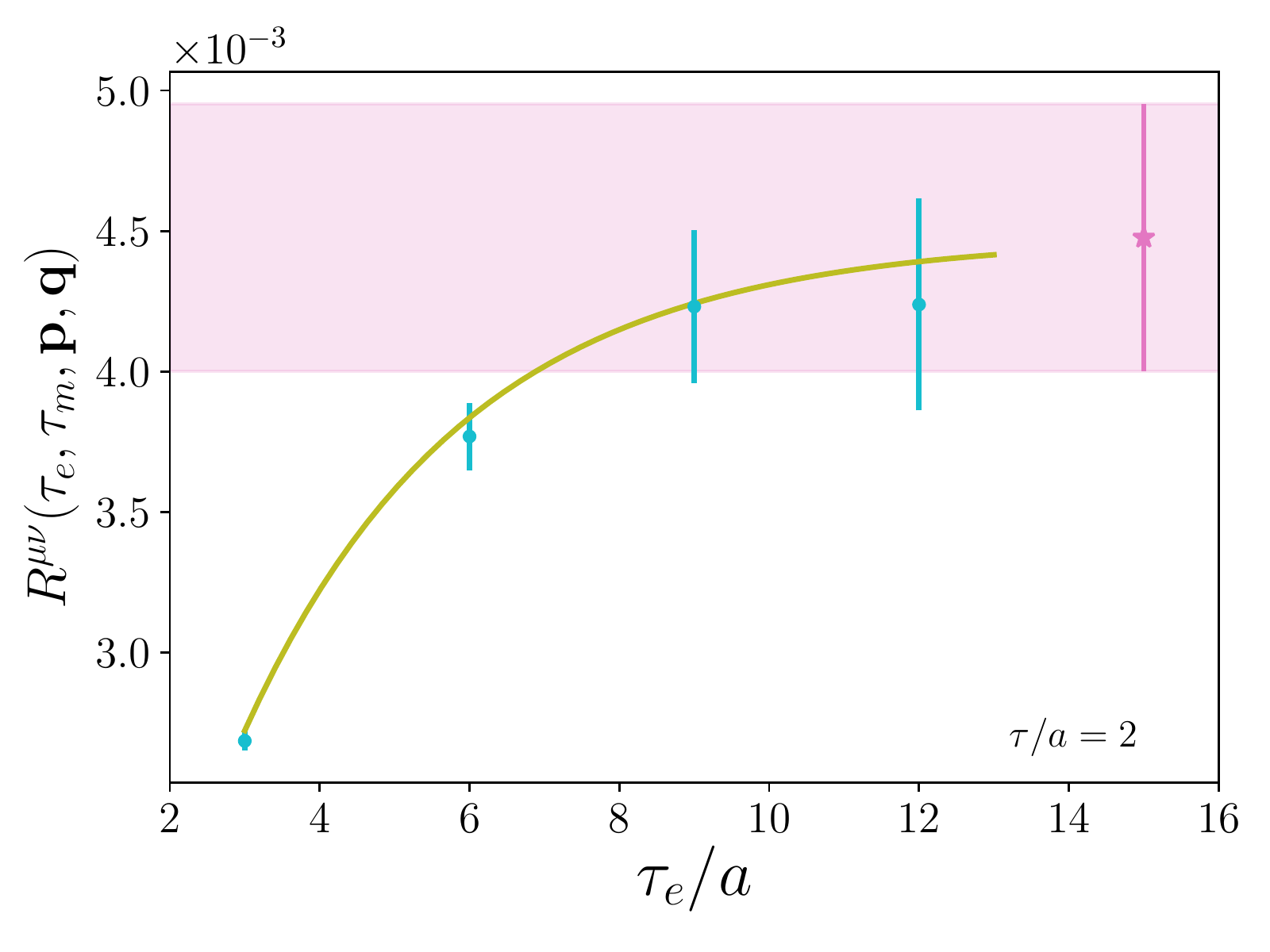}
    \includegraphics[width=0.49\textwidth]{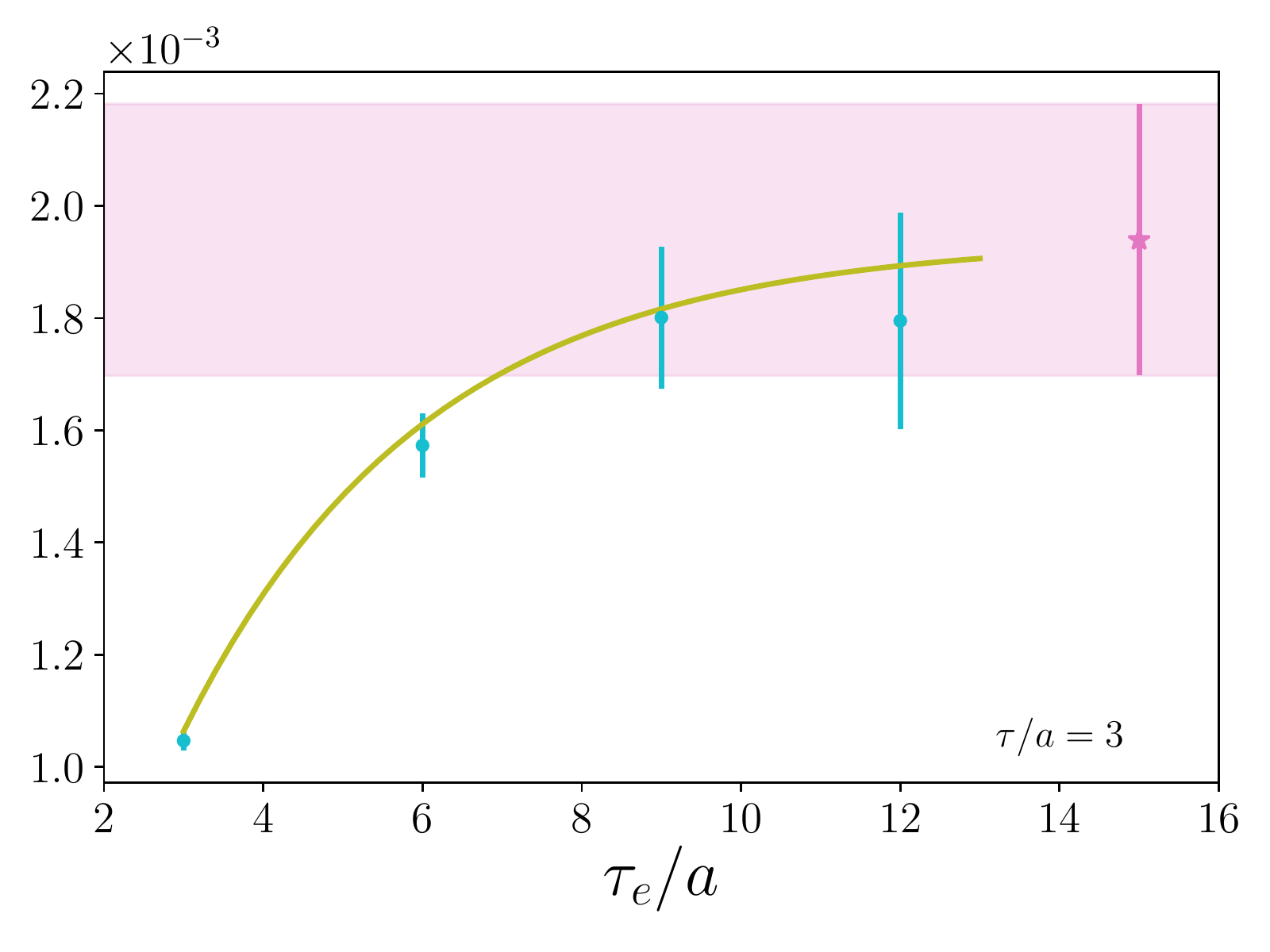}
    \caption{Examining the excited state contamination of the time-momentum representation data. Numerical data for the three-point correlator was calculated using a sequential source located at time slices $\tau_e/a=3,6,9,12$. By studying the data as a function of $\tau_e/a$ for fixed $\tau/a$, the excited state contamination may be quantified. The data obeys Eq.~\eqref{eq:excited_state}. The shaded band and star indicate the extrapolated values. However, in the analysis discussed below, fixed $\tau_e/a=9$ was taken. Thus the extracted values of the second and fourth moments contain residual excited state contamination.}
    \label{fig:excited_state_xi4}
\end{figure}

Excited state contamination in $R^{[\mu\nu]}(\tau_e,\tau_m,\mathbf{p},\mathbf{q};a)$ may be studied by varying the sequential source insertion time, $\tau_e$. Excited state contamination is expected to appear as
\begin{equation}
\label{eq:excited_state}
R^{[\mu\nu]}(\tau_e,\tau_m,\mathbf{p},\mathbf{q};a)=R_{(0)}^{[\mu\nu]}(\tau,\mathbf{p},\mathbf{q};a)\bigg(1+A e^{-\Delta E\tau_e}\bigg)\,,
\end{equation}
where $R_{(0)}^{[\mu\nu]}(\tau,\mathbf{p},\mathbf{q};a)$ is the ratio in the large Euclidean time limit, and $A$ and $\Delta E$ encode the excited state contamination. This matrix element was studied for $\tau_e/a=3,6,9,12$, thus allowing for a fit to the above form. The $\tau_e$ dependence for fixed $\tau/a$ is shown in Fig.~\ref{fig:excited_state_xi4}. From this figure, it can be seen that the data at fixed Euclidean time $\tau_e/a=9$ agrees within errors with the extrapolated value using Eq.~\eqref{eq:excited_state}. For this preliminary analysis, data was analysed at fixed Euclidean time, with $\tau_e/a=9$ taken for this analysis. As a result, it is expected that the resulting $\expval{\xi^4}$ determination contains a systematic error from this approximation. Taking central differences between data at $\tau_e/a=9$ and the extrapolated value is a $\sim 10\%$ effect. This uncertainty is smaller than the current statistical error. While this approach is sufficient for this initial analysis, with more statistical samples, it is clear that more sophisticated analysis methods should be used to further reduce the systematic error arising from excited state contamination.

\subsection{Data Analysis}

At the time of performing the second moment analysis, it was unclear which of the two approaches described in Sec.~\ref{sec:analysis} would lead to a smaller uncertainty and thus the fourth moment analysis used the momentum-space approach. Since the conference, it has become clear that while statistical uncertainties obtained from the two approaches are comparable, systematic errors are less well controlled in the momentum-space analysis, at least for the second moment~\cite{Detmold:2021qln}. 
However, since this exploratory study of $\expval{\xi^4}$ only consists of one lattice spacing and two heavy quark masses, no study of the systematic errors was performed.

Data was converted to momentum space by performing a discrete Fourier transform
\begin{equation}
V^{[\mu\nu]}(p,q;a)=a\sum_{\tau=-\tau_\text{max}}^{\tau_\text{max}} e^{iq_4\tau}R^{[\mu\nu]}(\tau,\mathbf{p},\mathbf{q};a)\,.
\end{equation}
where $\tau_\text{max}/a=11$.
While data at non-zero $\tau$ are guaranteed to have a well-defined continuum limit, $\tau = 0$ data contain additional
UV divergences arising from the mixing of the current-current operator with lower-dimensional operators. After
performing the Fourier transform, this divergence will appear as an additive shift in the numerical data. Thus a single subtraction is first performed at fixed lattice spacing:
\begin{equation}
V_\text{sub}^{[\mu\nu]}(p,q;a)=V^{[\mu\nu]}(p,q;a)-V^{[\mu\nu]}(p,(\mathbf{q},q_{4,\text{sub}});a)\,.
\end{equation}
In this analysis, the subtraction points was taken to be $q_{4,\text{sub}}=4\gev$.

\begin{figure}
    \centering
    \includegraphics[scale=0.45]{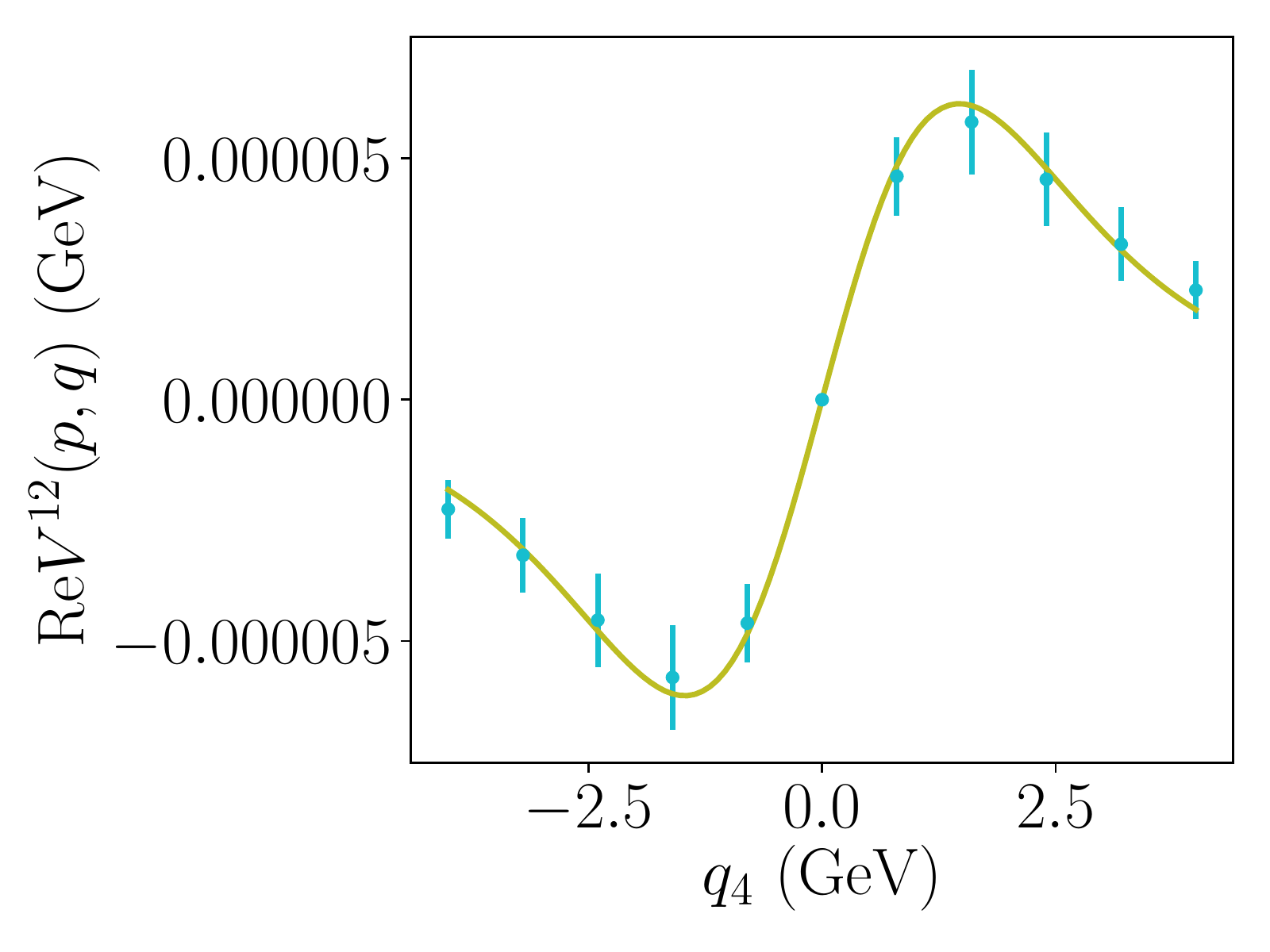}
    \includegraphics[scale=0.45]{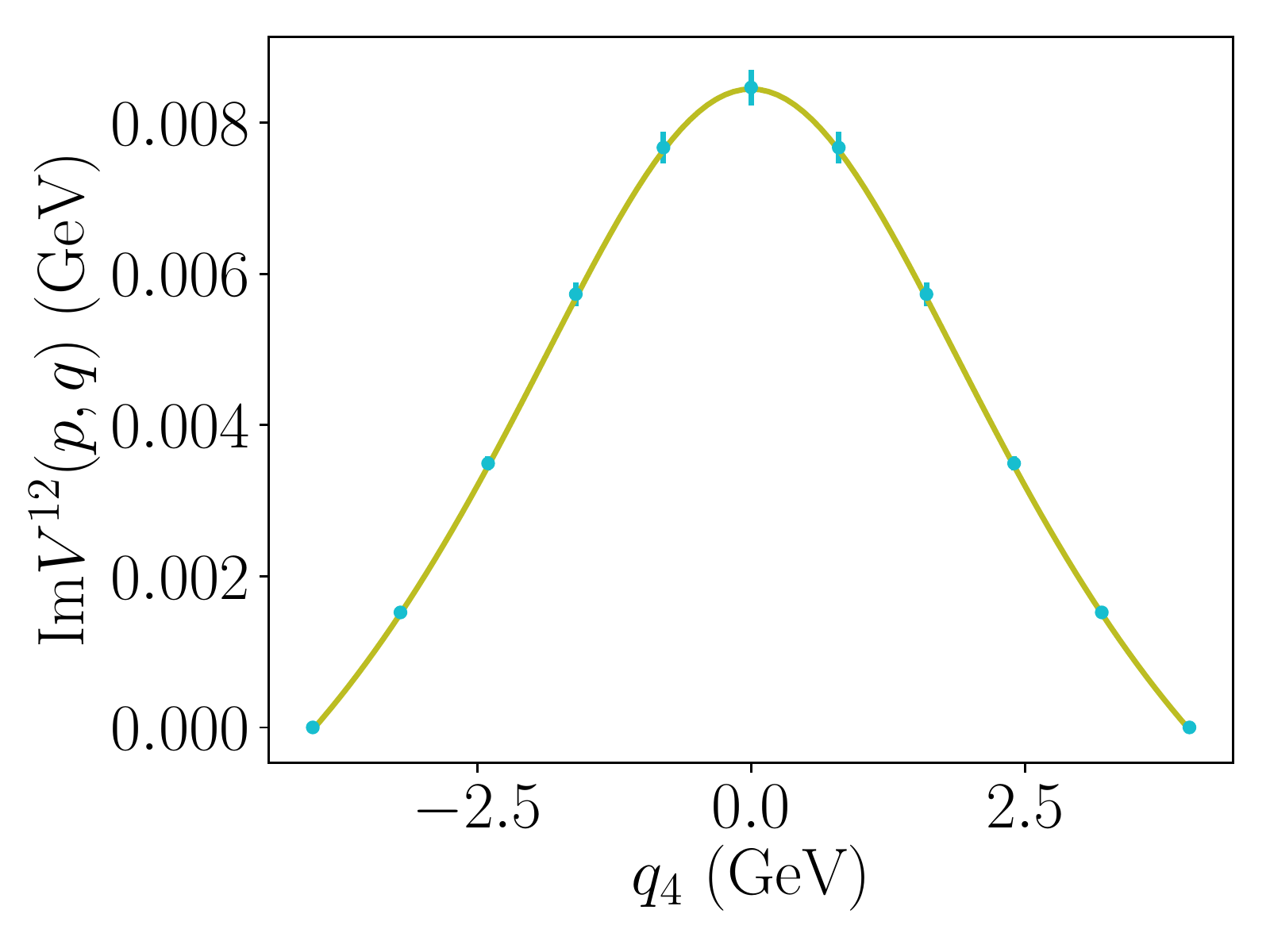}
    \caption{Fitting the real and imaginary parts of the numerical data  to the HOPE formula for $|\hat{\mathbf{p}}|=1$. This data was used in the second moment analysis described above. At these kinematics, $\expval{\xi^4}$ is negligible, and thus this data is used to extract $f_\pi$, $m_\Psi$ and $\expval{\xi^2}$.}
    \label{fig:mom_space_low_mom}
\end{figure}

The resulting lattice data is fit to the continuum form of the HOPE formula given in Eq.~\eqref{eq:OPE}. As a result, it is anticipated that lattice artifacts and higher-twist effects will be present in the determined parameters. The sensitivity to the fourth moment may be improved by utilizing the low-momentum data used in the second moment analysis to extract the parameters $f_\pi$, $m_\Psi$ and $\expval{\xi^2}$. Note that it is not expected that the size of the lattice artifacts are the same in the time-momentum representation and momentum-space data, although the fitted parameters should agree when extrapolated to the continuum limit. The fits to lattice correlators at low momentum are shown in Fig.~\ref{fig:mom_space_low_mom}. Utilizing these best fit parameters for the analysis of the high momentum ($|\mathbf{p}|\sim 1.3\gev$) data, it is possible to perform a single parameter fit to extract the fourth moment contribution from the two heavy quark masses. Both data sets comprised the observable calculated on 3500 gauge field configurations, and resulted in a fit value of $\expval{\xi^4}\sim 0.15$, with approximately a fifty percent statistical error.
It is important to note that these fit results are exploratory. At this point, no effort has been made to quantify systematic errors arising from lattice artifacts or higher-twist effects. In order to study these effects, more numerical data at a range of different lattice spacings, momenta, and heavy quark masses will be required.

To the best of the authors' knowledge, this constitutes the first attempt to determine the fourth Mellin moment of the pion LCDA from LQCD. There are however a number of predictions of the quantity from phenomenological approaches to QCD, which tend to predict a fourth moment of $\expval{\xi^4}(\mu\sim2\gev)\in(0.1,0.15)$~\cite{Bakulev:2002uc,Praszalowicz:2001wy,Chang:2013nia}. 

\begin{figure}
    \centering
    \includegraphics[scale=0.45]{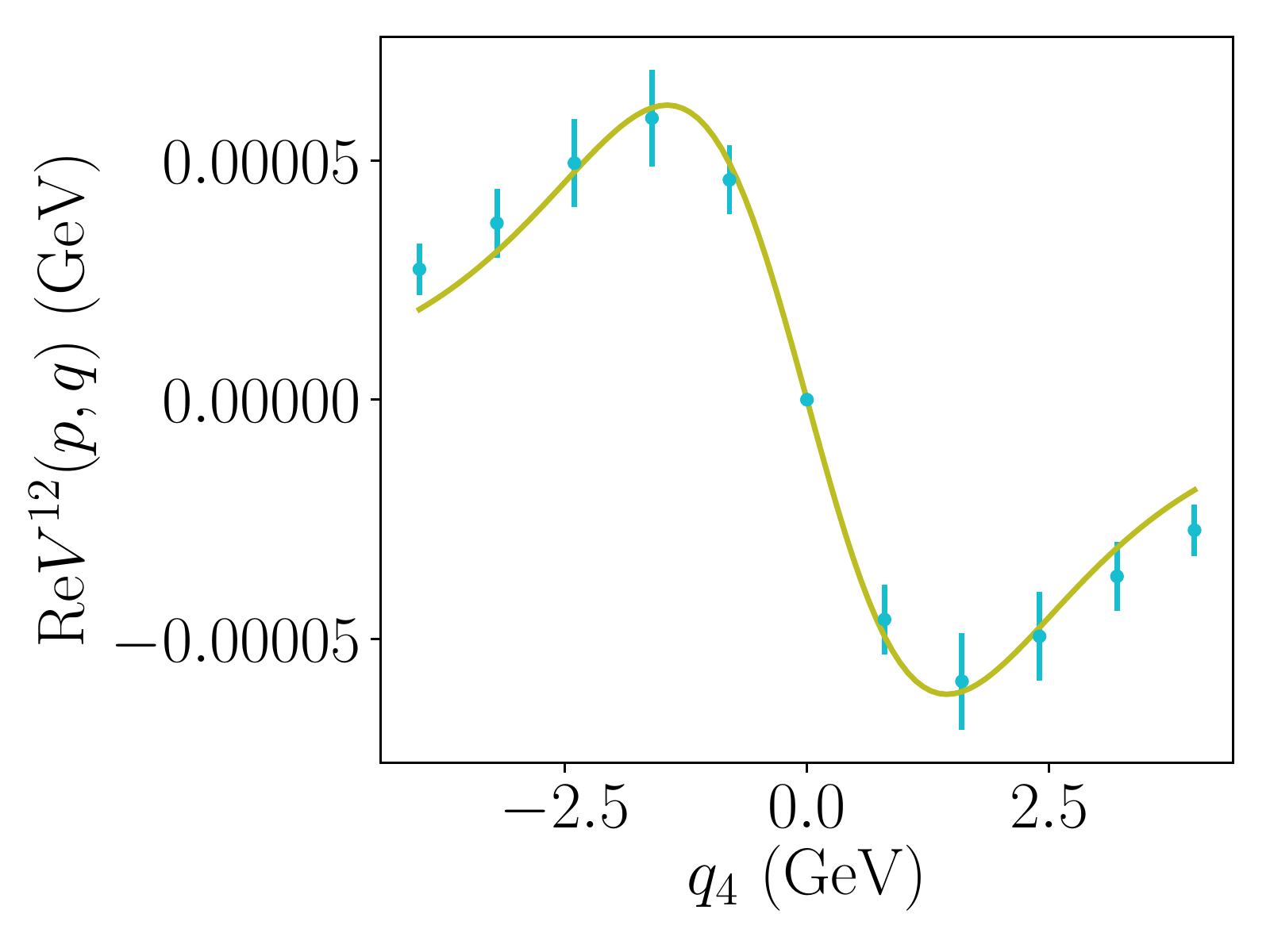}
    \includegraphics[scale=0.45]{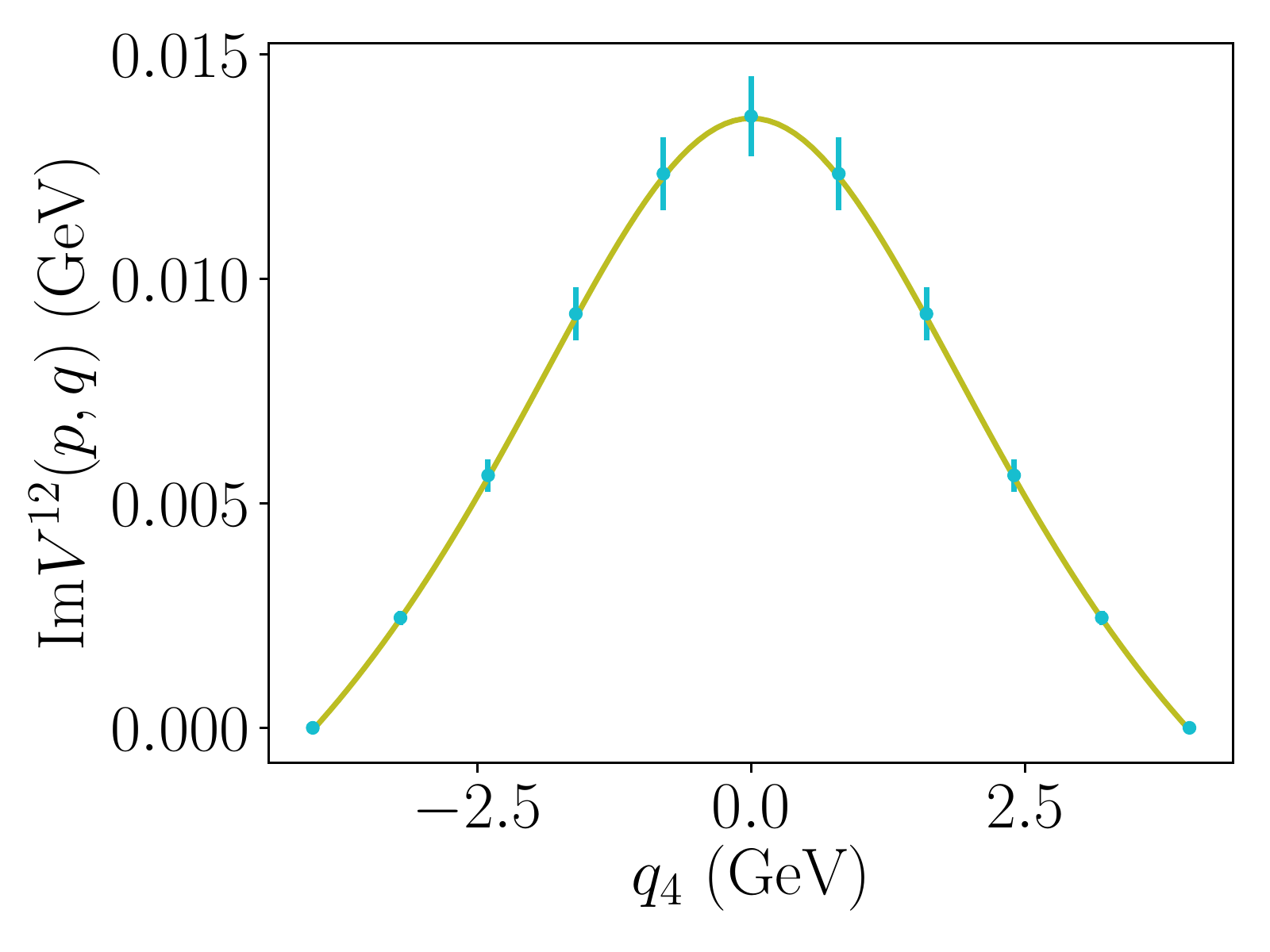}
    \caption{Fitting the real and imaginary parts of the numerical data to the HOPE formula for $|\hat{\mathbf{p}}|=2$. By using the fit parameters from the low-momentum data, the high momentum data is used to constrain $\expval{\xi^4}$.}
    \label{fig:}
\end{figure}

\section{Conclusion and outlook}
\label{sec:conclusion}
This article presents the progress on the first numerical study of the HOPE method to extract the second and fourth Mellin moments of the pion LCDA. The HOPE method utilizes a quenched fictitious heavy-quark species which enables further control over higher-twist contributions than that provided by large momentum alone. In this article, the second moment is determined by analysing the time-momentum representation of the lattice correlator, and is found to be
\begin{equation*}
  \langle \xi^2 \rangle(\mu=2\gev) = 0.210 \pm 0.036\,.
\end{equation*}
where the uncertainty in this result is dominated by systematic effects, in particular from higher-twist terms and the continuum extrapolation.  Additionally, this article presents an exploratory investigation of the fourth Mellin moment. The extracted value contains (at this point) significant systematic errors from excited state contamination, lattice artifacts and higher-twist effects. With additional numerical data, it will become possible to perform a more complete analysis of this quantity to ensure sufficient control of systematic errors.

Taken together, these results demonstrate the viability of the HOPE approach to determine moments of light-cone quantities with comparable statistical precision to that seen in results from other methods~\cite{Del_Debbio_2003,Arthur_2011,Bali:2019dqc}. This paves the way for further investigations of the pion LCDA, including dynamical studies of the second moment using the HOPE method and a determination of higher Mellin moments. The success of this approach for the LCDA also suggests that the HOPE method can be applied to the study of other light cone quantities, including (for example) the kaon LCDA and pion PDF and helicity PDF.

\section*{Acknowledgements}
The authors thank ASRock Rack Inc.~for their support of the construction of an
Intel Knights Landing cluster at National Yang Ming Chiao Tung
University, where the numerical calculations were performed.  Help
from Balint Joo in tuning Chroma is acknowledged.
We thank M. Endres for providing the ensembles of gauge field configurations used in this work.
  CJDL and RJP are supported by the Taiwanese MoST Grant
No.~109-2112-M-009-006-MY3 and MoST Grant No.~109-2811-M-009-516. 
The work of IK is partially supported by the MEXT as ``Program for
Promoting Researches on the Supercomputer Fugaku'' (Simulation for
basic science: from fundamental laws of particles to creation of
nuclei) and JICFuS.
YZ is supported by the U.S.~Department of Energy, Office of Science, Office of Nuclear Physics, contract no.~DEAC02-06CH11357.
WD and AVG acknowledge support from the U.S.~Department of Energy (DOE) grant DE-SC0011090. WD is also supported within the framework of the TMD Topical Collaboration of the U.S.~DOE Office of Nuclear Physics, and by the SciDAC4 award DE-SC0018121.
WD is also supported in part by the National Science Foundation
under Cooperative Agreement PHY-2019786 (The NSF AI Institute for Artificial Intelligence and Fundamental Interactions, http://iaifi.org/).
SM is partly supported by the LANL LDRD program.

\addcontentsline{toc}{chapter}{Bibliography} 
\bibliographystyle{jhep.bst} 

\bibliography{refs} 
\end{document}